\documentclass[11pt]{article}
\usepackage{amsmath,amssymb,mathrsfs,color,epsfig,cite}
\usepackage{graphicx}
\usepackage{subfigure}
\usepackage{setspace}

\usepackage{amsfonts}

\newcommand{\hoch}[1]{$\, ^{#1}$}

\makeatletter
\@addtoreset{equation}{section}
\makeatother

\newcommand{\be}{\begin{equation}}
\newcommand{\ee}{\end{equation}}
\newcommand{\bea}{\setlength\arraycolsep{2pt} \begin{eqnarray}}
\newcommand{\eea}{\end{eqnarray}}
\newcommand{\nn}{\nonumber}

\def\ft#1#2{{\textstyle{\frac{\scriptstyle #1}{\scriptstyle #2} } }}
\def\fft#1#2{{\frac{#1}{#2}}}

\def\0{{\sst{(0)}}}
\def\1{{\sst{(1)}}}
\def\2{{\sst{(2)}}}
\def\3{{\sst{(3)}}}
\def\4{{\sst{(4)}}}
\def\5{{\sst{(5)}}}
\def\6{{\sst{(6)}}}
\def\7{{\sst{(7)}}}
\def\8{{\sst{(8)}}}
\def\sst#1{{\scriptscriptstyle #1}}
\def\oneone{\rlap 1\mkern4mu{\rm l}}

\def\del{{\partial}}

\def\ed{{{\rm\bf d}}}

\def\cA{{{\cal A}}}
\def\cF{{{\cal F}}}
\def\cG{{{\cal G}}}
\def\cB{{{\cal B}}}

\def\tA{{{\widetilde A}}}

\def\cj{{{\cal J}}}
\def\tc{{{\tilde c}}}
\def\ibox{{\square^{-1}}}
\def\iboxsq{{\square^{-2}}}
\def\iboxcub{{\square^{-3}}}

\def\cG{{{\cal G}}}

\def\cH{{{\cal H}}}
\def\im{{{\rm i\,}}}

\def\wtd{\widetilde}

\def\cA{{{\cal A}}}

\def\bx{{{\bf x}}}
\def\by{{{\bf y}}}

\def\aa{{{\scriptstyle{(a)}}}}
\def\sone{{{\scriptstyle{(1)}}}}
\def\stwo{{{\scriptstyle{(2)}}}}
\def\sthree{{{\scriptstyle{(3)}}}}

\begin{document}

\begin{flushright}
\hfill { MI-HET-783
}\\
\end{flushright}

\begin{center}
{\large {\bf BRST Symmetry and the Convolutional Double Copy
 }}

\vspace{25pt}
{\large Mahdi Godazgar$^1$, 
              C.N. Pope$^{2,3}$, A. Saha$^2$ and Haoyu Zhang$^{2}$}

\vspace{25pt}

\hoch{1}{\it School of Mathematical Sciences, Queen Mary University of London,\\
   Mile End Road, E1 4NS, UK}

\hoch{2}{\it George P. \& Cynthia Woods Mitchell  Institute
for Fundamental Physics and Astronomy,\\
Texas A\&M University, College Station, TX 77843, USA}

\hoch{3}{\it DAMTP, Centre for Mathematical Sciences,
 Cambridge University,\\  Wilberforce Road, Cambridge CB3 OWA, UK}

\vspace{20pt}



\end{center}




\begin{abstract}

Motivated by the results of Anastasiou et al., 
we consider the convolutional double copy for BRST and anti-BRST covariant formulations of gravitational and gauge theories in more detail.  We give a general BRST and anti-BRST invariant formulation of linearised $\mathcal{N}=0$ supergravity using superspace methods and show how this may be obtained from the square of linearised Yang-Mills theories. We demonstrate this relation for the Schwarzschild black hole and the ten-dimensional black string solution as two concrete examples.
\end{abstract}

\vfill
{\scriptsize
m.godazgar@qmul.ac.uk, pope@physics.tamu.edu, aritrasaha@physics.tamu.edu,
zhanghaoyu@tamu.edu.}
\pagebreak

\tableofcontents
\addtocontents{toc}{\protect\setcounter{tocdepth}{2}}

\section{Introduction}

The emergence in recent years of a powerful ``double copy'' relation between 
gravitational and gauge theory scattering amplitudes 
\cite{Bern:2008qj, Bern:2010ue}, mirroring earlier such relations found in 
string theory \cite{Kawai:1985xq}, has led to a wide variety of applications, 
including the ability to calculate higher-loop supergravity amplitudes and 
in gravitational wave physics (see \cite{Bern:2022wqg} and references cited 
therein).  Given the by now well-established double copy relation for 
scattering amplitudes, a clear question worth pursuing is whether this 
relation is reflected at the level of classical solutions 
(see \cite{Kosower:2022yvp} and references cited therein).  Most of the 
explicit such relationships that have been found, such that the Kerr-Schild 
double copy \cite{Monteiro:2014cda} or the Weyl double copy 
\cite{Luna:2018dpt}, work for a special class of gravitational solutions.  
However, the convolutional double copy 
\cite{YMSq, Anastasiou:2017nsz, duffbrst} posits a square relationship 
between gauge and gravitational theories at the level of the fields, 
albeit at the linearised level.  We shall be focusing on this approach in 
this paper.

The idea of a convolutional double copy goes back to \cite{Borsten:2013bp}, 
where the global U-duality groups are analysed. However, more significantly 
it is shown in \cite{YMSq} that the convolutional product of gauge theory 
symmetries reproduces the symmetries of the gravitational theory at the 
linear level. That there may be a convolutional square relation between 
gauge theory and gravitational fields in position space is not so 
surprising if we recall the double copy relation for scattering amplitudes, 
which are derived in momentum space.  In \cite{duffbrst}, taking the 
low-energy limit of the bosonic string ($\mathcal{N}=0$ supergravity) as a 
concrete gravitational theory, it is found that in order to have a matching 
of the degrees of freedom in this theory, i.e.\ ($D^2-2D+2$), and the 
square of the Yang-Mills theory, i.e.\ $(D-1)^2$, as well as to be able to 
disentangle the dilaton and the trace of the metric, one ought to consider a 
BRST covariant formulation 
for each of the theories.  Thus, a set of ghost fields 
are introduced for the gravitational theory, as well as the gauge theories.  
This sets up a dynamical equivalence between gauge theories squared and 
gravitational theories.\footnote{The idea that BRST ghosts are needed 
in order to establish the double copy relations was presaged in the
work of Siegel in \cite{siegel1995}.}  
However, an important assumption in this 
construction is that all fields are sourced by non-trivial currents, which 
makes it possible formally to invert the d'Alembertian operator appearing 
in the equations of motion.  

In this paper, we revisit \cite{duffbrst} and consider the procedure outlined 
there more generally and in more detail.  In particular, we present a 
detailed study of a BRST and anti-BRST covariant formulation of the 
$\mathcal{N} = 0$ theory.  We show that the 2-form gauge invariance may 
be neatly treated using the superspace method developed in 
\cite{Baulieu:1983tg, Perry:1992yj}.  The BRST and anti-BRST Lagrangian is 
then constructed by adding to the original gauge-invariant Lagrangian a 
gauge-fixing contribution that is by construction BRST and anti-BRST 
invariant. One of the facts that we emphasise in this work is that because 
the inverse of the d'Alembertian operator may be used freely, this means 
that the source term is in principle ambiguous in the sense that there
is some freedom to choose which terms are taken to be contributing to 
the ${\cal N}=0$ supergravity sources and which terms 
are not. This is particularly important when dealing with explicit 
solutions.

In section \ref{sec:BRST}, we derive a general BRST and anti-BRST invariant 
formulation of the ${\mathcal{N}=0}$ theory and show how the action used 
in \cite{duffbrst} may be obtained as a special case of 
this general formulation.  In 
section \ref{sec:GravYM}, we reconsider in greater detail the convolutional 
relation studied in \cite{duffbrst}.  In contrast to \cite{duffbrst}, we do 
not constrain the gravitational field ans\"atze using the equations of 
motion, because as explained above terms may be absorbed into the source 
term, making such conditions arbitrary.  We rely solely on the BRST and 
anti-BRST transformations to constrain the ans\"atze.  We derive the 
double copy relation for the sources in section \ref{sec:FieldEqns}.  
Our results are broadly in agreement with those of \cite{duffbrst}.  We 
highlight the fact that there is freedom in choosing the coefficients in 
the ans\"atze.  In particular, the choice made by \cite{duffbrst} leads to 
a breakdown of the double copy relation when the Yang-Mills gauge-fixing 
parameter $\xi=1$.  This behaviour is an artefact of the choice made in 
\cite{duffbrst} and there is another choice for which this behaviour is 
not observed and moreover the equation for the gravitational source term 
is simpler.  

  In section \ref{sec:BH}, we demonstrate the convolutional 
double copy relation for two classical solutions: the Schwarzschild black 
hole and the ten-dimensional black string solution.  Black holes have 
been considered in this context in 
\cite{Cardoso:2016ngt, Cardoso:2016amd, LopesCardoso:2018xes, point}.  
However, here, we derive the gravitational fields from the ans\"atze 
derived in section \ref{sec:GravYM}.  Interestingly, in order to
obtain a convolutional double copy description of the black string,
we need to use Yang-Mills fields that lie in an $SU(2)$ subgroup of
the gauge group.  This contrasts with the double copy description of
the black hole, where it suffices to consider Yang-Mills fields in a
$U(1)$ subgroup of the gauge group.  We also discuss the BPS limit
of the black string, for which it turns out that only an abelian $U(1)$
subgroup of the Yang-Mills gauge group is required.

  After conclusions in section \ref{sec:conc}, we include also an
appendix, where we summarise some basic results about the convolution
product that is employed in the double-copy relations throughout
the paper.  We also discuss the notion of the convolution inverse, 
which plays a role in the double-copy relations, and in particular
we highlight the fact that care must be taken ensure that the
convolution inverse actually exists for the functions one is using.

\section{BRST Symmetry and ${\cal N}=0$ Supergravity}\label{sec:BRST}

\subsection{The linearised theory}

  The theory that arises as the low-energy limit of the bosonic string is
sometimes referred to as ``${\cal N}=0$ supergravity.''  Its field
content comprises the metric $g_{\mu\nu}$, the dilaton $\varphi$ and the
antisymmetric tensor potential $B_{\mu\nu}$, with the equations
of motion following from the Lagrangian
\bea
{\cal L}= \sqrt{-g}\, \Big[R - \ft12 (\del\varphi)^2 - 
\ft1{12} e^{-a\varphi}\, H^2\Big]\,,
\eea
where $H=dB$ and
\bea
a=\sqrt{\fft{8}{D-2}}\,,
\eea
with $D$ being the dimension of the spacetime.  

    Our discussion will be restricted to the linearisation of the theory 
around a Minkowski spacetime background, with the metric written as
\bea
g_{\mu\nu}= \eta_{\mu\nu} + h_{\mu\nu}.
\eea
At this order, the Lagrangian is simply given by
\bea
{\cal L}(h_{\mu\nu},\varphi, B_{\mu\nu})
 = -\ft12 h^{\mu\nu}\, G^{\rm lin}_{\mu\nu} -\ft12 (\del\varphi)^2 
   -\ft1{12} H^{\mu\nu\rho}\, H_{\mu\nu\rho}\,,\label{hphiBlag}
\eea
where
\bea
G^{\rm lin}_{\mu\nu}= -\ft12 \square h_{\mu\nu} +
  \del_\sigma\del_{(\mu} h_{\nu)}{}^\sigma -
   \ft12\del_\mu\del_\nu h -\ft12\del_\rho\del_\sigma h^{\rho\sigma}\,
\eta_{\mu\nu} +\ft12 \square h\, \eta_{\mu\nu}
\eea
is the linearised Einstein tensor, $h=\eta^{\mu\nu}\, h_{\mu\nu}$, and
indices are raised and lowered using the Minkowski metric.
This Lagrangian is invariant under linearised diffeomorphisms and 
2-form gauge transformations, namely under
\bea
\delta h_{\mu\nu}= 2\del_{(\mu} \xi_{\nu)}\,,\qquad
\delta B_{\mu\nu}= 2\del_{[\mu} \lambda_{\nu]}\,,\qquad
\delta\varphi=0\,.\label{gaugetrans}
\eea

\subsection{General formulation invariant under BRST and anti-BRST}

  In a BRST formulation, the diffeomorphism and 2-form gauge invariances
are handled by the introduction of diffeomorphism ghosts $c_\mu$ and 
 $\bar c_\mu$, and 2-form gauge invariance ghosts $d_\mu$ and $\bar d_\mu$.
The implementation of the BRST formulation for the 2-form gauge symmetry is
a little more complicated than that for the diffeomorphism invariance,
because of the ``ghosts for ghosts'' phenomenon, stemming from the fact that
gauge parameters $\lambda_\mu$ in eqn (\ref{gaugetrans}) that are themselves
of the form $\lambda_\mu=\del_\mu\lambda$ do not contribute.  The
upshot from this is that in addition to the ghost $d_\mu$ and antighost 
$\bar d_\mu$ fields one must also introduce fields $d$, $\bar d$ and $\eta$.
We shall discuss this in detail below, but first we consider the ghost
$c_\mu$ and antighost $\bar c_\mu$ in the gravity sector.

  In the gravity sector, the BRST transformations in the linearised
theory are as follows:
\bea
Q h_{\mu\nu}= 2\del_{(\mu} c_{\nu)}\,,\qquad Q c_\mu=0\,,\qquad
\qquad Q \bar c_\mu =W_\mu\,,\qquad Q W_\mu=0\,.\label{BRSTc}
\eea
There are also anti-BRST transformations
\bea
\bar Q h_{\mu\nu}= 2\del_{(\mu} \bar c_{\nu)}\,,\qquad \bar Q \bar c_\mu=0\,,\qquad
\qquad \bar Q c_\mu = -W_\mu\,,\qquad \bar Q W_\mu=0\,.\label{aBRSTc}
\eea
Later, $W_\mu$ will be seen to be the gauge-fixing functional for the
diffeomorphism symmetry.  The BRST and anti-BRST transformations 
can be seen to anticommute when acting on any of the fields, i.e.\ 
$\{Q,\bar Q\}=0$.  Note that this is an off-shell statement: no equations
of motion are involved.

  The BRST and anti-BRST description of the 2-form gauge symmetry requires a
lengthier discussion, which follows in the next subsection.

\subsection{BRST and anti-BRST for the 2-form gauge invariance}
\label{BRSTgensec}

   Our discussion here is based upon the approach of \cite{Baulieu:1983tg, Perry:1992yj}.  
  We define a superspace exterior derivative operator\footnote{We are using the boldface symbol $\ed$ to denote exterior derivatives in order to avoid 
confusion with the ghost number 2 field $d$.}
\bea
\hat\ed = \ed + s + \bar s\,,
\eea
where 
\bea
\ed= \ed x^\mu\,\fft{\del}{\del x^\mu}\,,\qquad
s= \ed\theta\, \fft{\del}{\del\theta}\,,\qquad
\bar s= \ed\bar\theta\, \fft{\del}{\del\bar\theta}\,,
\eea
We also define a super 2-form $\cB$ as
\bea
\cB = B -d_\mu\, \ed x^\mu\wedge \ed\theta - 
    \bar d_\mu\, \ed x^\mu\wedge \ed\bar\theta -
   d\, \ed\theta\wedge\ed\theta  -
  \bar d\,\ed\bar\theta\wedge\ed\bar\theta -
   \eta\, \ed\theta\wedge \ed\bar\theta\,,
\eea
where $B$ is just the usual bosonic 2-form,
\bea
B=\ft12 B_{\mu\nu}\,\ed x^\mu\wedge \ed x^\nu\,,
\eea
and  $d_\mu, {\bar d}_\mu, d, {\bar d}$ and $\eta$ are ghost fields with ghost numbers $+1, -1, +2, -2$ and $0$ respectively.

 Fields are taken to be functions of $x^\mu$, $\theta$ and
$\bar\theta$, with theta expansions of the form
\bea
B_{\mu\nu}(x,\theta,\bar\theta)= B_{\mu\nu}(x) + Q B_{\mu\nu}(x)\,\theta
+ \bar Q B_{\mu\nu}(x)\, \bar\theta + \cdots\,,
\eea
etc.  Here, $Q$ denotes the BRST operator and $\bar Q$ the
anti-BRST operator.  In what follows, it will be assumed that after
exterior derivatives are taken, $\theta$ and $\bar\theta$ are set to zero.

  We now find that $\cH\equiv \hat\ed \cB$ is given by
\bea
\cH &=&  H+ (\ft12 Q B_{\mu\nu} -\del_{[\mu} d_{\nu]})\,
  \ed x^\mu\wedge \ed x^\nu\wedge \ed\theta +
  (\ft12\bar Q B_{\mu\nu} -\del_{[\mu} \bar d_{\nu]})\,
  \ed x^\mu\wedge \ed x^\nu\wedge \ed\bar\theta \nn\\
&&-
(\del_\mu d -Q d_\mu)\, \ed x^\mu\wedge \ed\theta\wedge\ed\theta -
(\del_\mu \bar d - \bar Q \bar d_\mu)\, 
     \ed x^\mu\wedge \ed\bar\theta\wedge\ed\bar\theta\nn\\
&& -
(\del_\mu \eta - Q\bar d_\mu - \bar Q d_\mu)\, 
\ed x^\mu\wedge \ed\theta\wedge\ed\bar\theta 
- Q d\,\ed\theta\wedge\ed\theta\wedge\ed\theta -
\bar Q \bar d\, \ed\bar\theta\wedge\ed\bar\theta\wedge\ed\bar\theta 
\nn\\
&&-
(Q\eta + \bar Q d)\, \ed\theta\wedge\ed\theta\wedge\ed\bar\theta -
(\bar Q \eta +Q \bar d)\, \ed\theta\wedge\ed\bar\theta\wedge\ed\bar\theta
\,,
\eea
where
\bea
H= \ed B= 
\ft16 H_{\mu\nu\rho}\, \ed x^\mu\wedge \ed x^\nu\wedge \ed x^\rho
\eea
is the purely bosonic 3-form field strength.  

Requiring that all the
components of $\cH$ with projections in the $\theta$ and $\bar\theta$
directions must vanish gives the equations
\bea
Q B_{\mu\nu} &=& 2\del_{[\mu} d_{\nu]}\,,\qquad Q d_\mu =\del_\mu d\,,\qquad
Q d=0\,,\nn\\
\bar Q B_{\mu\nu} &=& 2\del_{[\mu} \bar d_{\nu]}\,,
\qquad \bar Q \bar d_\mu =\del_\mu\bar d\,,\qquad
\bar Q \bar d=0\,,\label{Qtrans1}
\eea
together with
\bea
Q \bar d + \bar Q \eta =0\,,\qquad \bar Q d + Q\eta=0\,,\qquad
\del_\mu\eta - Q\bar d_\mu -\bar Q d_\mu=0\,.\label{mixed}
\eea 
These last three equations can be rewritten as separated BRST and 
anti-BRST transformations by introducing auxiliary ghost 
fields $\tau$ and $\bar\tau$ and an auxiliary bosonic field $K_\mu$, such
that eqns (\ref{mixed})  become\footnote{Note that in \cite{Bonora:2007hw}, the last equation in (\ref{mixed}) is written in terms of
two distinct auxiliary fields $B_\mu$ and $\bar B_\mu$ as 
\bea
Q \bar d_\mu=B_\mu\,,\qquad \bar Q d_\mu=-\bar B_\mu\nn
\eea
together with the constraint $B_\mu-\bar B_\mu -\del_\mu\eta=0$ that
would then have to be imposed by hand.  Instead, here, we are using just
the single auxiliary field $K_\mu$, in terms of which
the fields of \cite{Bonora:2007hw} are given by $B_\mu=K_\mu +\ft12\del_\mu\eta$ and
$\bar B_\mu = K_\mu -\ft12 \del_\mu\eta$. The advantage of our
current approach using just $K_\mu$ is that there is no need to 
impose any constraint equation.}
\bea
Q\bar d&=& -\bar\tau\,,\qquad \bar Q\eta= \bar\tau\,,\qquad\qquad
\bar Q d = -\tau\,,\qquad Q\eta= \tau\,,\label{taueqns}\\
Q \bar d_\mu &=& K_\mu +\ft12 \del_\mu\eta\,,\qquad
\bar Q d_\mu = -K_\mu +\ft12 \del_\mu\eta\,,\label{Keqns}
\eea
where $K_\mu$ is for now arbitrary.  Eventually, it will 
acquire the interpretation of being an auxiliary field whose algebraic equation
follows after the introduction of a gauge-fixing and ghost action.
It follows also that
\bea
Q\tau &=&0\,,\qquad  \bar Q \tau=0\,,\qquad Q\bar\tau=0\,,\qquad 
\bar Q\bar\tau=0\,,\nn\\
Q K_\mu &=& -\ft12\del_\mu\tau\,,\qquad \bar Q K_\mu =\ft12\del_\mu\bar\tau\,.
\label{Qtrans4}
\eea
Again, $\tau$ and $\bar\tau$ are arbitrary at this stage, and will later 
be seen to be auxiliary fields that are determined by the form of the gauge-fixing and ghost action.

\subsection{General class of BRST and anti-BRST invariant theories}
\label{theorygensec}

  We can now construct a general class of theories by considering a Lagrangian
\bea
{\cal L}= {\cal L}(h_{\mu\nu},\varphi,B_{\mu\nu}) + \hat{\cal L} \,,
\label{Ltot}
\eea
where ${\cal L}(h_{\mu\nu},\varphi,B_{\mu\nu})$ is the set of
(gauge-invariant) kinetic terms for $B_{\mu\nu}$, $h_{\mu\nu}$ and $\varphi$,
and
\bea
\hat{\cal L}= Q\bar Q\, X=-\bar Q Q\, X\,,\label{hatlag2}
\eea
where $X$ is the most general quadratic function built from the fields
$h_{\mu\nu}$, $\varphi$, $B_{\mu\nu}$, $c_\mu$, $\bar c_\mu$, 
$d_\mu$, $\bar d_\mu$, $d$,
$\bar d$, $\eta$, consistent with having vanishing ghost number
 and being Lorentz invariant.
We also impose the restriction that
the Lagrangian should not contain any terms with more than two 
derivatives. The most general such function can be taken
to be
\bea
X&=& \beta_1\, \bar d d + \beta_2\, \bar d^\mu d_\mu +
 \beta_3\, B^{\mu\nu} B_{\mu\nu} + \gamma_1\, \bar c^\mu c_\mu +
\gamma_2\, h^{\mu\nu} h_{\mu\nu} + \gamma_3\, h^2 \nn\\
&& +
\beta_4\, \bar d^\mu  c_\mu + \bar\beta_4\, \bar c^\mu d_\mu +
\beta_5\, h\eta + \beta_6\, h\varphi\,.\label{genX2}
\eea
$\hat{\cal L}$ constructed by this means will comprise the
gauge-fixing and ghost terms in the full Lagrangian ${\cal L}$ given in
eqn (\ref{Ltot}).

Using eqns (\ref{Qtrans1}), (\ref{taueqns}), (\ref{Keqns}) 
 and (\ref{Qtrans4}), we then find from
(\ref{hatlag2}) that
\bea
\hat {\cal L} &=& \beta_1\, \bar\tau \tau +
\beta_2\, [\bar\tau\del^\mu d_\mu -\del^\mu\bar d_\mu \tau
   -\bar d\square d + K^\mu K_\mu-\ft14 (\del\eta)^2] \nn\\
&&+
  4\beta_3\, (\bar d^\mu\square d_\mu -\bar d_\mu \del^\mu\del^\nu d_\nu
+ B^{\mu\nu}\, \del_\mu K_\nu) 
+ \gamma_1\, W^\mu W_\mu \nn\\
&&
+ \gamma_2(8\del^{(\mu} c^{\nu)}\del_\mu\bar c_\nu
+ 4 h^{\mu\nu} \del_\mu W_\nu)  +
  \gamma_3\,(8 \del^\mu c_\mu\del^\nu \bar c_\nu + 4 h\del^\mu W_\mu)\nn\\
&&
+ \beta_4\, (\bar\tau \del^\mu c_\mu + W^\mu K_\mu +\ft12 W^\mu\,\del_\mu\eta)
+
 \bar\beta_4 (-\del^\mu\bar c_\mu\tau + W^\mu K_\mu-
          \ft12 W^\mu\,\del_\mu\eta)\nn\\
&& -
2\beta_5\, (W^\mu\del_\mu\eta + \del^\mu\bar c_\mu \tau
 + \bar\tau \del^\mu c_\mu) - 2\beta_6\, W^\mu \del_\mu\varphi\,.
\label{hatlag}
\eea
Recall that in this formulation we have $\hat{\cal L}= Q\bar Q X =
-\bar Q Q X$ off-shell and with no need for imposing any constraint, 
and so $\hat{\cal L}$
is manifestly invariant under both $Q$ and $\bar Q$ transformations.

  Solving for the auxiliary fields $W_\mu$ and $K_\mu$, we find
\bea
W_\mu &=& \fft1{\Delta}\,\Big\{ -8\beta_2\, (\gamma_2\, \del^\nu h_{\nu\mu}
  + \gamma_3\, \del_\mu h +\ft12\beta_6\, \del_\mu\varphi) +
  4\beta_3\, (\beta_4+\bar\beta_4)\, \del^\nu B_{\nu\mu}\nn\\
&&\qquad +
  \beta_2\, (\beta_4-\bar\beta_4 - 4 \beta_5)\del_\mu\eta
\Big\}\,,\\
K_\mu&=& \fft1{\Delta}\,\Big\{4(\beta_4+\bar\beta_4)\,
(\gamma_2\, \del^\nu h_{\nu\mu} + \gamma_3\, \del_\mu h
   +\ft12\beta_6\, \del_\mu\varphi) -
    8\beta_3\, \gamma_1\, \del^\nu B_{\nu\mu}\nn\\
&&\qquad
   -\ft12(\beta_4+\bar\beta_4)(\beta_4-\bar\beta_4-4\beta_5)\del_\mu\eta\Big\}\,,
\label{eqaux1}
\eea
where
\bea
\Delta = (\beta_4 + \bar\beta_4)^2 -4\beta_2\, \gamma_1\,.
\eea

   We can also solve the algebraic equations of motion for the auxiliary
fields $\tau$ and $\bar\tau$ that follow from the Lagrangian (\ref{hatlag}),
finding
\bea
\tau &=& -\fft1{\beta_1}\, \Big[ \beta_2\, \del^\mu d_\mu + 
 (\beta_4-2\beta_5)\del^\mu c_\mu\Big]\,,\nn\\
\bar\tau &=& \fft1{\beta_1}\, \Big[ \beta_2\,\del^\mu\bar d_\mu +
(\bar\beta_4+2\beta_5) \del^\mu\bar c_\mu\Big]\,.
\label{eqaux2}
\eea

\subsection{Specialisation to Ref.~\cite{duffbrst}}

   It is instructive to compare the general expressions we have obtained
in sections \ref{BRSTgensec} and \ref{theorygensec} with the form of the 
BRST transformations and 
Lagrangian discussed in \cite{duffbrst}.  Scaling up the Lagrangian in
\cite{duffbrst} by a factor of 2 in order to agree with our normalisation,
it becomes
\bea
{\cal{L}}' = {\cal{L}}(h_{\mu\nu},\varphi, B_{\mu\nu}) + \hat{\cal{L}'}\,,
\label{Dlag}
\eea
where ${\cal L}(h_{\mu\nu},\varphi, B_{\mu\nu})$ is given by eqn (\ref{hphiBlag})
and
\bea
\hat{\cal{L}'} &=& -2 \bar c^\mu\square c_\mu - 2 \bar d^\mu \square d_\mu 
+\Big(2 - \fft{2m_d}{\xi_d}\Big)\, \bar d_\mu \del^\mu\del^\nu d_\nu +
  2 m_d\, \bar d \square d + \fft1{\xi_h}\, \big(\del^\nu h_{\nu\mu} -
\ft12 \del_\mu h\big)^2\nn\\
&& + \fft1{\xi_B}\, \big(\del^\nu B_{\nu\mu} +
\del_\mu\eta'\big)^2\,.\label{Dhatlag}
\eea
Note that in matching the BRST transformations in \cite{duffbrst} with
our transformations in section \ref{BRSTgensec} we can see that the
normalisations of $c_\mu$, $d_\mu$ and $d$ (and their conjugates) are the
same, but we must introduce a scaling factor that relates our $\eta$ ghost
and the corresponding $\eta'$ ghost in \cite{duffbrst}. From a matching of the BRST transformations we deduce that our parameters in the $\hat{\cal{L}'}$ Lagrangian \eqref{Dhatlag}, and the relation between $\eta$ and $\eta'$, must be chosen so that
\bea
\beta_4 &=& -\bar\beta_4 = 2\beta_5\,,\qquad \beta_6=0\,,\qquad
\gamma_3=-\ft12 \gamma_2\,,\nn\\
\gamma_1 &=& 2\gamma_2\, \xi_h\,,\qquad \beta_2=2\beta_3\, \xi_B\,,\qquad 
\beta_1=-\beta_2\, \xi_d\,, \qquad \eta'= \ft12 \xi_B\, \eta\,.
\eea
Consistency also requires that $m_d = \ft12 \xi_B$.

After substituting the equations of motion of auxiliary fields in $\hat{\cal{L}}$, and comparing the Lagrangians (i.e. $\hat{\cal{L}}$ and $\hat{\cal{L}'}$), we find that we must take $\beta_3=\gamma_2=-\ft12$.
Thus, to summarise, we find that the BRST formulation in \cite{duffbrst}
corresponds to the specialisation of our general formulation in which
\bea
\beta_1&=& \xi_B\, \xi_d\,,\qquad
\beta_2=-\xi_B\,,\qquad \beta_3=-\ft12\,,\qquad
\beta_4=-\bar\beta_4=2\beta_5\,,\qquad\beta_6=0\,,\nn\\
\gamma_1 &=& -\xi_h\,,\qquad \gamma_2=-\ft12\,,\qquad \gamma_3=\ft14\,.
\label{coeffs}
\eea
Furthermore, in order for the formulation in \cite{duffbrst} to exhibit
both BRST and anti-BRST invariance, it is necessary that the parameters
$m_d$ and $\xi_B$ obey the relation
\bea
m_d=\ft12 \xi_B\,.\label{mxi}
\eea
Note that with the choices in eqns (\ref{coeffs}) for the coefficients,
we have
\bea
W_\mu= \fft1{\xi_h}\, (\del^\nu h_{\nu\mu} -\ft12 \del_\mu h)\,,\qquad 
K_\mu= \fft1{\xi_B}\, \del^\nu B_{\nu\mu}\,.\label{WKcanonical}
\eea
Thus $Q\bar c_\mu= W_\mu$ and $Q\bar d_\mu= K_\mu +\ft12 \del_\mu\eta=
\fft1{\xi_B}\, (\del^\nu B_{\nu\mu} + \del\eta')$ can be seen to have the standard forms for
De Donder and 2-form Lorenz gauge-fixing functionals.

We finish this section by summarising the BRST and anti-BRST transformations of the fields in this specialisation.  These will be needed in section \ref{sec:BRSTdc}.  The BRST transformations are given by
\begin{gather}
 Q h_{\mu\nu} = 2\del_{(\mu} c_{\nu)}\,,\qquad Q c_\mu=0\,, \qquad Q \bar c_\mu =\fft1{\xi_h}\, (\del^\nu h_{\nu\mu} -\ft12 \del_\mu h)\,, \qquad Q \varphi = 0\,, \notag \\[2mm] 
 Q B_{\mu\nu} = 2\del_{[\mu} d_{\nu]}\,,\qquad Q d_\mu =\del_\mu d\,,\qquad Q d=0\,,\nn \\[2mm] 
 Q \bar d_\mu = \fft{1}{\xi_B}\, (\del^\nu B_{\nu\mu} + \del_\mu\eta')\,, \qquad Q \bar d = \fft{1}{\xi_d}\, \del^\mu \bar d_\mu\,, \qquad Q\eta'= \fft{m_d}{\xi_d}\, \del^\mu d_\mu.
 \label{summ:BRST}
\end{gather}
The anti-BRST transformations are:
\begin{gather}
\bar Q h_{\mu\nu}= 2\del_{(\mu} \bar c_{\nu)}\,,\qquad \bar Q \bar c_\mu=0\,, \qquad \bar Q c_\mu = -\fft1{\xi_h}\, (\del^\nu h_{\nu\mu} -\ft12 \del_\mu h)\,, \qquad \bar Q \varphi = 0\,, \notag \\[2mm]
\bar Q B_{\mu\nu} = 2\del_{[\mu} \bar d_{\nu]}\,, \qquad \bar Q \bar d_\mu =\del_\mu\bar d\,,\qquad \bar Q \bar d=0\,, \nn \\[2mm]
\bar Q d_\mu = - \fft{1}{\xi_B}\, (\del^\nu B_{\nu\mu} - \del_\mu\eta'), \qquad \bar Q d = - \fft{1}{\xi_d}\, \del^\mu d_\mu\,, \qquad \bar Q\eta'=- \fft{m_d}{\xi_d}\, \del^\mu \bar d_\mu.
\label{summ:aBRST}
\end{gather}

\section{${\cal N}=0$ Supergravity from the Square of Yang-Mills} \label{sec:GravYM}

As outlined in the introduction, there are various realisations of the double copy idea at the level of solutions and theories, each with its own features and challenges.  Here, we will be focusing on the convolutional double copy of \cite{YMSq}.  The basic idea here is that in momentum space, the double copy relation satisfied by the scattering amplitudes is one of a product, i.e.\ a gravitational amplitude is roughly obtained by taking the product of two Yang-Mills amplitudes.  Therefore, at the level of fields, expressed in position space, it is natural to expect that the double copy relation will be one of convolution. More precisely, at the linearised level, one expects a relation of the form
\begin{equation} \label{convprod}
 H_{\mu \nu} = A_{\mu}^a \star \phi_{a \tilde{b}}^{-1} \star \tilde{A}_\nu^{\tilde{b}} \equiv A_{\mu}^a \circ \tilde{A}_\nu^{\tilde{b}},
\end{equation}
where $H_{\mu \nu}$ is a composite field containing the metric perturbation 
$h_{\mu \nu}$, two-form field $B_{\mu \nu}$ and the dilaton $\varphi$, 
and $\phi_{a\tilde b}$ is the spectator bi-adjoint scalar field \cite{Hodges:2011wm, Cachazo:2013iea}, which is used to soak up the potentially different gauge indices. $\star$ denotes a convolution, while $\circ$ denotes a convolutive product with the pseudo-inverse of the bi-adjoint scalar suppressed, for brevity.  (See appendix A for a more detailed discussion
of the convolution product.) 
Thus, in this way one can view the gravitational theory as 
the square of a gauge theory. Moreover, the linearised gravitational symmetries may be derived from those of the gauge theory.

Off-shell, the gravitational theory has $(D^2-2D+2)$ degrees of 
freedom: $D(D-1)/2$ from the metric, $(D-1)(D-2)/2$ from the 2-form and 1 
from the scalar field.  In contrast, each one of the gauge theories 
has $(D-1)$  degrees of freedom, which come from the gauge field. Taking the convolutive product of gauge theory fields then gives $(D-1)^2$ degrees of freedom, which is one short of the gravitational theory. This problem is resolved in \cite{duffbrst} by including ghost fields to take care of the gauge symmetry, leading to BRST covariant theories. We discussed the BRST covariant formulation 
of the gravitational theory in section \ref{sec:BRST}, and we shall
review the BRST formulation of Yang-Mills theory in \ref{sec:YM}.  
The gravitational theory, then, has fields 
$(h_{\mu \nu}, B_{\mu \nu}, \varphi, c_\mu, \bar{c}_\mu, d_{\mu}, 
\bar{d}_\mu, d, \bar{d}, \eta)$ and the two Yang-Mills theories 
have fields $(A_{\mu}, c, \bar{c})$ and $(\tA_{\mu}, \tc, \bar{\tc})$ 
(see section \ref{sec:YM}).  The convolutive product of the ghost fields of the Yang-Mills theories gives 4 degrees of freedom, which accounts for the degrees of freedom in $(d, \bar{d}, \eta)$ and a combination of $h$ and $\varphi$, which we can simply think of as accounting for $\varphi$.  The convolutive product of the ghost sector and the gauge fields then contributes $4(D-1)$ degrees of freedom, which account for the gravitational ghost fields $(c_\mu, \bar{c}_\mu, d_{\mu}, \bar{d}_\mu)$.  Finally, now, the $(D-1)^2$ degrees of freedom coming from the convolutive product of the gauge fields accounts for the degrees of freedom in $h_{\mu \nu}$ and $B_{\mu \nu}.$  In addition, the introduction of the ghost fields allows one to disentangle the degrees of the freedom in $h$ and $\varphi$.

Crucially, the formalism developed in \cite{duffbrst} requires the presence of sources so that one can work with the non-local operator $\Box^{-1}$,
\begin{equation}
 \Box^{-1} \Box = \Box\, \Box^{-1} = \oneone.
\end{equation}
Thus, this construction cannot deal with source-free 
vacuum solutions, such as gravitational waves.

\subsection{BRST formulation of linearised Yang-Mills}\label{sec:YM}

  Following the proposal in \cite{duffbrst} for deriving the linearised
${\cal N}=0$ supergravity theory as the convolution product of
two copies of linearised Yang-Mills, we begin by reviewing the BRST
formulation for the linearised Yang-Mills theories.    Thus we begin 
with the Lagrangian 
\bea
{\cal L}_A = {\rm tr\,}\Big( -\ft14 F^{\mu\nu}\, F_{\mu\nu} +
   \fft1{\xi}\, (\del^\mu A_\mu)^2 -\bar c\,\square c\Big)\,,\label{YMlag}
\eea
where at the linearised level we can just take 
$F_{\mu\nu}=2\del_{[\mu} A_{\nu]}$\,.  The equations of motion for the
Yang-Mills and ghost fields are, after introducing sources, given by
\bea
\square A_\mu -\fft{\xi+1}{\xi}\, \del_\mu\del^\nu A_\nu = j_\mu(A)\,,
\qquad \square c^\alpha = j^\alpha(c)\,,
\eea
where, following \cite{duffbrst}, the ghost $c$ and antighost $\bar c$ are
grouped into an $OSp(2)$ doublet $c^\alpha$, with
\bea
c^1=c\,,\qquad c^2= \bar c\,,\qquad c_1=\bar c\,,\qquad c_2=-c\,.
\eea

   The constant $\xi$ parameterises a family of Lorenz gauge fixings.  
Making momentum space expansions
\bea
A_\mu =\int d^4 p\, \cA_\mu(p)\, e^{\im p\cdot x}\,,\qquad
j_\mu = \int d^4 p\, \cj_\mu(p)\, e^{\im p\cdot x}\,,
\eea
we have
\bea
-p^2\, \cA_\mu + \fft{\xi+1}{\xi}\, p_\mu\, p^\nu\, \cA_\nu =\cj_\mu\,.
\label{momeom}
\eea
The propagator $\cG_{\mu\nu}$ has the property that
\bea
\cA_\mu = \cG_{\mu\nu}\, \cj^\nu\,.\label{Asol2}
\eea
Making the ansatz that
\bea
\cG_{\mu\nu}= \alpha\, \eta_{\mu\nu} + \beta\, p_\mu\, p_\nu\,,
\eea
we can plug $\cj_\mu$ from eqn (\ref{momeom}) into eqn (\ref{Asol2}), and
hence solve for $\alpha$ and $\beta$.  The result is that the propagator is
given by
\bea
\cG_{\mu\nu}= -\fft{\eta_{\mu\nu} - (\xi+1)\, \fft{p_\mu\,p_\nu}{p^2}}{p^2}\,.
\label{propagator}
\eea
Taking $\xi=-1$ corresponds to the Feynman-'t Hooft
propagator.  Note that if we  make no gauge fixing (i.e.\ take $\xi=\infty$),
the propagator becomes singular, as expected.

   The theory (\ref{YMlag}) is invariant under the BRST transformations
\bea
Q A_\mu=\del_\mu c\,,\qquad Q c=0\,,
 \qquad Q\bar c= \fft1{\xi}\, \del^\mu A_\mu\,.\label{YMBRST}
\eea
It is also invariant under the anti-BRST transformations
\bea
\bar Q A_\mu=\del_\mu \bar c\,,\qquad \bar Q \bar c=0\,,
 \qquad \bar Q c= -\fft1{\xi}\, \del^\mu A_\mu\,.\label{YMaBRST}
\eea

\subsection{BRST, anti-BRST and the convolutional double copy}
\label{sec:BRSTdc}

  Following \cite{duffbrst}, we begin by writing ans\"atze
for the fields of linearised
${\cal N}=0$ supergravity as the convolution product of two
copies of linearised Yang-Mills (one untilded and one tilded), as follows:
\bea
\varphi &=& A^\rho\circ \tA_\rho + \alpha_1\,c^\alpha\circ \tc_\alpha
            +\alpha_2\, \ibox\, (\del A\circ\del\tA)\,,\label{phians}\\
h_{\mu\nu} &=& A_{(\mu}\circ \tA_{\nu)} + a_1\, \ibox\,\del_\mu\del_\nu\,
(A^\rho\circ\tA_\rho) + 
  a_2\, \ibox\, \del_\mu\del_\nu (c^\alpha \circ \tc_\alpha)\nn\\
&& +a_3\, \ibox\, \Big(\del A\circ \del_{(\mu} \tA_{\nu)} + 
         \del_{(\mu} A_{\nu)}\circ \del\tA \Big) + 
      a_4\, \iboxsq\,\del_\mu\del_\nu (\del A\circ\del\tA) \nn\\
&& +\eta_{\mu\nu}\,\Big(b_1\, A^\rho\circ \tA_\rho + 
     b_2\,c^\alpha\circ \tc_\alpha
            + b_3\, \ibox\, (\del A\circ\del\tA) \Big)\,,\label{hans}\\
B_{\mu\nu} &=& A_{[\mu} \circ \tA_{\nu]} + 
  \gamma\,\ibox\, (\del A\circ \del_{[\mu} \tA_{\nu]} 
    -\del_{[\mu}A_{\nu]}\circ \del \tA)\,,\label{Bans}
\eea
where $\del A$ denotes $\del^\mu A_\mu$ and similarly for $\del \tA$.  Note that these ans\"atze are the
most general possible, compatible with the conditions of bi-linearity in
the two sets of Yang-Mills fields, and the preservation of the $OSp(2)$
symmetry of the ghost sectors.  It is useful to note that since
$c^\alpha\circ \tc_\alpha= c\circ \bar\tc - \bar c\circ\tc$, we have
from (\ref{YMBRST}) and (\ref{YMaBRST}) that
\bea
Q\, (c^\alpha\circ \tc_\alpha)= 
   -\fft1{\xi}\, (c\circ \del \tA +\del A\circ \tc)\,,\qquad
\bar Q\, (c^\alpha\circ \tc_\alpha)=
   -\fft1{\xi}\, (\bar c\circ \del \tA +\del A\circ \bar\tc)\,.
\eea

   Note that the Yang-Mills gauge and ghost fields
in the ans\"atze (\ref{phians}), (\ref{hans}) and (\ref{Bans}) should
be thought of as
carrying gauge-group indices that are suppressed and traced over
in the convolution products.  

  Constraints on the constant coefficients in the ans\"atze follow by
requiring that the fields $h_{\mu\nu}$, 
$\varphi$ and $B_{\mu\nu}$ should satisfy their BRST and anti-BRST 
transformation rules, as discussed in section 2, as a consequence of the
BRST and anti-BRST transformation rules (\ref{YMBRST}) and (\ref{YMaBRST}).  For definiteness, we shall assume in this
discussion that the parameters in the gauge-fixing and ghost Lagrangian
$\hat{\cal L}$ in eqn (\ref{hatlag}) have been specialised as in eqns
(\ref{coeffs}) and (\ref{mxi}), so that the gauge-fixing functionals
$W_\mu$ and $K_\mu$ take the forms given in eqns (\ref{WKcanonical}).  The BRST and anti-BRST 
transformations in this case are summarised in \eqref{summ:BRST} and \eqref{summ:aBRST}, respectively.

  Starting with the dilaton field $\varphi$, the requirement that 
$Q\varphi=0$ implies
\bea
1-\fft{\alpha_1}{\xi} + \alpha_2=0\,.\label{Qphi}
\eea
We obtain the identical condition from considering instead $\bar Q\varphi=0$.

    From $Q h_{\mu\nu}=2\del_{(\mu}\, c_{\nu)}$, we deduce that
\bea
b_1 -\fft{b_2}{\xi} + b_3=0\label{eq1}
\eea
and 
\bea
c_\mu = \ft12 (1+a_3)\, (c\circ \tA_\mu + A_\mu\circ \tc ) +
  \ft12 (a_1-\fft{a_2}{\xi} + a_3 + a_4)\, \ibox\del_\mu
   (c\circ\del\tA + \del A\circ\tc)\,.
\eea
Using instead $\bar Q h_{\mu\nu}=2\del_{(\mu}\, \bar c_{\nu)}$ gives
the same condition (\ref{eq1}) and also
\bea
\bar c_\mu = \ft12 (1+a_3)\, (\bar c\circ \tA_\mu + A_\mu\circ \bar\tc ) +
  \ft12 (a_1-\fft{a_2}{\xi} + a_3 + a_4)\, \ibox\del_\mu
   (\bar c\circ\del\tA + \del A\circ \bar\tc)\,.
\eea
So in fact $\bar c_\mu$ is precisely what one would get by conjugating
the expression for $c_\mu$.

  From $Q\bar c_\mu= \xi_h^{-1}\, (\del^\nu h_{\nu\mu}-\ft12 \del_\mu h)$
we then find
\bea
\fft{1+a_3}{2\xi} &=& \fft{1+a_3}{2\xi_h}\,,\label{eq2}\\
1+ a_1 -\fft{a_2}{\xi} + 2 a_3 + a_4 &=&\fft{a_2 + (2-D) b_2}{\xi_h}\,,
   \label{eq3}\\
\fft1{\xi}\, (a_1-\fft{a_2}{\xi} + a_3 + a_4) &=&
\fft1{2\xi_h}\, [a_4 + (2-D) b_3]\,,\label{eq4}\\
0 &=& a_1 - 1 + (2-D) b_1\,.\label{eq5}
\eea
Thus eqn (\ref{eq2}) implies $\xi_h=\xi$.  Note that eqns (\ref{eq3}),
(\ref{eq4}) and (\ref{eq5}) also imply eqn (\ref{eq1}).

  The upshot is that in the $\varphi$ and $h_{\mu\nu}$ sectors, we may
think of $\alpha_1$, $a_1$, $a_2$, $a_3$ and $a_4$ as being
freely specifiable, with $\alpha_2$, $b_1$, $b_2$ and $b_3$ then being
determined by eqns (\ref{Qphi}), (\ref{eq3}), (\ref{eq4}) and (\ref{eq5}).

   From the ansatz for the $B_{\mu\nu}$ field in eqn (\ref{Bans}),
we can read off from the BRST transformation $QB_{\mu\nu}= 2\del_{[\mu}\,d_{\nu]}$ that
\bea
d_\mu= \ft12(\gamma+1)\, (c\circ \tA_\mu- A_\mu\circ \tc) +
  \beta\ibox\, \del_\mu(c\circ\del\tA - \del A\circ\tc)\,,\label{dmu0}
\eea
where $\beta$ is an as-yet undetermined constant.  Considering instead
$\bar QB_{\mu\nu}= 2\del_{[\mu}\,\bar d_{\nu]}$, we deduce
\bea
\bar d_\mu= \ft12(\gamma+1)\, (\bar c\circ \tA_\mu- A_\mu\circ \bar\tc) +
  \bar\beta\ibox\, \del_\mu(\bar c\circ\del\tA - \del A\circ\bar \tc)
\,,\label{bardmu0}
\eea
where $\bar\beta$ is an (a priori different) as-yet undetermined constant.

 Plugging $d_\mu$ given in eqn (\ref{dmu0}) into 
$Q d_\mu=\del_\mu d$ implies that
\bea
d= -(\gamma+1+2\beta)\, c\circ\tc\,.\label{dres}
\eea
Similarly, from $\bar Q \bar d_\mu=\del_\mu \bar d$ we obtain
\bea
\bar d= -(\gamma+1+2\bar\beta)\, \bar c\circ\bar \tc\,.\label{bardres}
\eea

Plugging $\bar d_\mu$ into 
$Q \bar d_\mu = \xi_B^{-1}\, (\del^\nu B_{\nu\mu} + \del_\mu\eta')$
implies that we must have 
\bea
\xi_B=\xi\,,\label{xiBres}
\eea
and also gives\footnote{Note that it is the rescaled field $\eta'=\ft12\xi_B\, \eta$ 
that appears in
eqn (\ref{Dhatlag}), and not the field $\eta$ that is defined in
section \ref{BRSTgensec}.}
\bea
\eta' = -\ft12(\gamma+1+2\bar \beta)\,  \xi_B\,
 (c\circ\bar{\tc} +\bar c\circ\tc)
\,.\label{etares}
\eea

 From $\bar Q d_\mu = -\xi_B^{-1}\, (\del^\nu B_{\nu\mu} - \del_\mu\eta')$
we again learn that $\xi_B=\xi$, and also we find 
\bea
\eta' = -\ft12 (\gamma+1+2\beta)\, \xi_B\, 
 (c\circ\bar{\tc} +\bar c\circ\tc)
\,.\label{etaresbar}
\eea
Comparing with eqn (\ref{etares}) therefore implies that the constants
$\beta$ and $\bar\beta$ are equal,
\bea
\bar\beta =\beta\,.
\eea

Plugging $\bar d$ from eqn (\ref{bardres}) 
into $Q \bar d = \xi_d^{-1}\, \del^\mu \bar d_\mu$ gives
\bea
\fft{\gamma+1+2\beta}{\xi} = \fft{\gamma+1+2\beta}{2\xi_d}\,,
\eea
and hence (assuming $(\gamma+1+2\beta)\ne0$) that
\bea
\xi_d= \ft12\xi\,.\label{xidres}
\eea
The same conclusion results from $\bar Q d= -\xi_d^{-1}\, 
 \del^\mu d_\mu$.

Plugging $\eta'$ given in eqn (\ref{etares}) into 
$Q\eta'= (m_d/\xi_d)\, \del^\mu d_\mu$ 
gives
\bea
(\gamma+1+2\beta)\, \Big(\fft{\xi_B}{\xi} - \fft{m_d}{\xi_d}\Big)=0\,,
\eea
and so (assuming $(\gamma+1+2\beta)\ne0$),
\bea
\fft{\xi_B}{\xi} = \fft{m_d}{\xi_d}\,.\label{mdeqn}
\eea

  Putting all the above together, we have
\bea
\xi_B=\xi\,,\qquad m_d=\xi_d = \ft12\xi\,.\label{ximres}
\eea
Note that these results imply, in particular, that the 
requirement in eqn (\ref{mxi}) which 
we previously found to be necessary in order
that $Q$ and $\bar Q$ anticommute is indeed obeyed.
Moreover, in the Kalb-Ramond sector, the coefficients $\beta$
and $\gamma$ are freely specifiable at this point.

\subsection{Field equations and the convolutional double copy} \label{sec:FieldEqns}

  Having extracted the various constraints on the coefficients in the
convolutional double copy ans\"atze (\ref{phians}), (\ref{hans}) and
(\ref{Bans}) that follow from requiring a consistency with the
BRST and anti-BRST transformations, we can now turn to a consideration
of the field equations.  For the fields $\varphi$, $h_{\mu\nu}$ and
$B_{\mu\nu}$, the equations of motion will be those following from
the Lagrangian ${\cal{L}}'$ given by eqns (\ref{Dlag}), (\ref{hphiBlag}) and (\ref{Dhatlag}), subject also to the condition (\ref{mxi}).  After
including source currents, the equations of motion are
\bea
\square\,\varphi &=& j(\varphi)\,,\label{phieom}\\
\square \, h_{\mu\nu} -\fft{\xi_h+2}{\xi_h}\, (2\del^\rho\del_{(\mu}
h_{\nu)\rho} -\del_\mu \del_\nu h) &=& j_{\mu\nu}(h)\,,\label{heom}\\
\square\, B_{\mu\nu} +\fft{\xi_B + 2}{\xi_B}\, 2\del^\rho \del_{[\mu}
B_{\nu]\rho} &=& j_{\mu\nu}(B)\,.\label{Beom}
\eea
We shall consider current sources for the ghost fields too, which thus obey
the equations of motion
\bea
\square \, c_\mu &=&j_\mu(c_\nu)\,,\qquad 
  \square \, \bar c_\mu =\bar j_\mu(\bar c_\nu)\,, \nn\\
\square\, d_\mu -\fft{\xi_d - m_d}{\xi_d}\, \del_\mu\del^\nu d_\nu 
  &=& j_\mu(d_\nu)\,,\qquad
\square\, \bar d_\mu -\fft{\xi_d - m_d}{\xi_d}\, \del_\mu\del^\nu \bar d_\nu
  = \bar j_\mu(\bar d_\nu)\,,\nn\\
\square \, d &=& j(d)\,,\qquad \square \,\bar d= \bar j(\bar d)\,,\qquad
\square\,\eta' = j(\eta')\,.\label{ghosteom}
\eea

  The idea now is to substitute the convolutional double copy ans\"atze into
the equations of motion.  This first of all provides a verification that
the equations for the fields of the ${\cal N}=0$ supergravity (together 
with its ghost fields) are indeed satisfied by virtue of the equations of
motion for the Yang-Mills fields (and their ghosts).  Additionally, one
can read off the expressions for the source currents of the
${\cal N}=0$ supergravity in terms of the source currents for the Yang-Mills
theories.

      It should be noted that by taking the divergence of the Yang-Mills
equation 
\bea
\square A_\mu -\fft{\xi+1}{\xi}\, \del_\mu \del A= j_\mu\label{YMeom}
\eea
we find $\square \del A= -\xi\, \del j$, where $\del j\equiv \del_\mu j^\mu$, and hence 
$\del A=-\xi\,\ibox\, \del j$.  Plugging this back into (\ref{YMeom})
allows us to solve for $A_\mu$ in terms of $j_\mu$, with
\bea
A_\mu=\ibox\, j_\mu - (\xi+1)\, \iboxsq\, \del_\mu \del j\,.\label{Asol}
\eea
 From this follow the results that
\bea
A^\rho\circ \tA_\rho = \iboxsq\, j^\rho\circ\tilde j_\rho
+ (\xi^2-1)\, \iboxcub\, \del j\circ\del\tilde j\,,\qquad
\del A\circ\del\tA = \xi^2\, \iboxsq\, \del j\circ\del\tilde j\,.
\label{Asqjsq}
\eea

   Consider first the equation of motion for the dilaton.  Plugging the
ansatz (\ref{phians}) into the equation of motion (\ref{phieom}) and using
the above results gives
\bea
\square\,\varphi=\ibox\, j^\rho\circ\tilde j_\rho + 
  \alpha_1\, \ibox\, j^\alpha(c)\circ \tilde j_\alpha(\tc) +
  [(1+\alpha_2)\, \xi^2 -1]\, \iboxsq\, \del j\circ\del\tilde j\,,
\eea
and hence\footnote{In \cite{duffbrst}, the choice $\alpha_1=1/\xi$ was made, which implies,
given the condition (\ref{Qphi}) found previously, that 
$\alpha_2=-1+1/\xi^2$ and hence the last term in (\ref{jphi}) vanishes.}
\bea
j(\varphi) = \ibox\, j^\rho\circ\tilde j_\rho +
  \alpha_1\, \ibox\, j^\alpha(c)\circ \tilde j_\alpha(\tc) +
  [(1+\alpha_2)\, \xi^2 -1]\, \iboxsq\, \del j\circ\del\tilde j\,.\label{jphi}
\eea

  Now consider the equation of motion for the graviton.  Plugging 
the ansatz (\ref{hans}) into the equation of motion (\ref{heom}), we
can obtain an expression for the graviton current (energy-momentum tensor):
\bea
j_{\mu\nu}(h)&=& \ibox\, j_{(\mu}\circ \tilde j_{\nu)} + 
   b_2\, \eta_{\mu\nu}\, \ibox\, j^\alpha(c)\circ\tilde j_\alpha(\tc) +
e_1\, \Big(\del A\circ \del_{(\mu} \tA_{\nu)} + 
   \del_{(\mu} A_{\nu)}\circ \del\tA\Big)
\nn\\
&&+ e_2\, \del_\mu\del_\nu (A^\rho\circ \tA_\rho) + 
  e_3\, \ibox\,\del_\mu\del_\nu\, (\del A \circ\del\tA) +
 e_4\, \eta_{\mu\nu}\, \del A\circ\del\tA \nn\\
&&+ 
 e_5\, \del_\mu\del_\nu\, (c^\alpha\circ\tc_\alpha)
  + b_1\, \eta_{\mu\nu}\, 
\ibox\, j^\rho\circ \tilde j_\rho\,,\label{hj1}
\eea
where the $e_i$ coefficients are given by\footnote{We have already used the equation $\xi_h=\xi$ that was derived previously
from the BRST and anti-BRST invariance conditions.}
\bea
e_1&=& -\fft{1+2 a_3}{\xi}\,,\qquad
e_2=\fft{[1+(D-2) b_1](\xi+2) -2 a_1}{\xi}\,,\nn\\
e_3 &=& \fft{(D-2)(2+\xi)\, b_3}{\xi} -
  \fft{(1+\xi)^2+2 a_4\, \xi}{\xi^2}\,,\qquad
e_4=\Big(1-\fft1{\xi^2}\Big)\, b_1 + b_3\,,\nn\\
e_5 &=& \fft{(D-2)(\xi+2)\, b_2 + 2 a_2}{\xi}\,.\label{edefs}
\eea
In view of the relation (\ref{Asol}), and the consequent equations
(\ref{Asqjsq}), the graviton current $j_{\mu\nu}(h)$ in eqn (\ref{hj1}) can be 
expressed entirely in terms of the Yang-Mills and Yang-Mills ghost
source currents, as
\bea
j_{\mu\nu}(h)&=& \ibox\, j_{(\mu}\circ \tilde j_{\nu)} +
   b_2\, \eta_{\mu\nu}\, \ibox\, j^\alpha(c)\circ\tilde j_\alpha(\tc) -
e_1\, \xi \,\iboxsq\, \Big(\del j\circ \del_{(\mu} \tilde j_{\nu)} +
   \del_{(\mu} j_{\nu)}\circ \del\tilde j\Big)
\nn\\
&&+ [2 e_1\, \xi(\xi+1) + e_2\, (\xi^2-1) + e_3\, \xi^2]\, \iboxcub\,
  \del_\mu\del_\nu\, (\del j\circ\del\tilde j) +
  e_2\, \iboxsq\,\del_\mu\del_\nu\, (j^\rho \circ\tilde j_\rho)\nn\\
&& +
 e_4\, \eta_{\mu\nu}\, \iboxsq\,(\del j\circ\del\tilde j) +
 e_5\, \iboxsq\,\del_\mu\del_\nu\, 
   (j^\alpha(c)\circ\tilde j_\alpha(\tc)) +
  b_1\, \eta_{\mu\nu}\,\ibox\, j^\rho\circ \tilde j_\rho \,.\label{hj2}
\eea

  For the Kalb-Ramond field $B_{\mu\nu}$ we find, plugging the ansatz (\ref{Bans})
into the equation of motion (\ref{Beom}) that the source current $j_{\mu\nu}(B)$
is given by\footnote{We have used here the already-derived result that $\xi_B=\xi$.}
\bea
j_{\mu\nu}(B)&=& \ibox\, j_{[\mu}\circ \tilde j_{\nu]} -
   \fft{2\gamma+1}{\xi}\, \big(\del A\circ \del_{[\mu} \tA_{\nu]} -
                  \del_{[\mu} A_{\nu]}\circ \del\tA\big)\,.\label{Bj1}
\eea
Written purely in terms of Yang-Mills source currents we therefore have
\bea
j_{\mu\nu}(B)&=& \ibox\, j_{[\mu}\circ \tilde j_{\nu]} +
  (2\gamma+1)\, \iboxsq\, \big(\del j\circ \del_{[\mu} \tilde j_{\nu]} -
         \del_{[\mu} j_{\nu]}\circ \del\tilde j\big)\,.\label{Bj2}
\eea

  In what has been done so far, we have simply given the general expressions
for the source currents $j(\varphi)$, $j_{\mu\nu}(h)$ and $j_{\mu\nu}(B)$
for $\varphi$, $h_{\mu\nu}$ and $B_{\mu\nu}$ in terms of the source 
currents of the Yang-Mills sectors.  This has not resulted in any further
restrictions on the coefficients in the original convolutional double
copy ans\"atze (\ref{phians}), (\ref{hans}) and (\ref{Bans}); the
only restrictions are those we already derived in section \ref{sec:BRSTdc}
by considering the BRST and anti-BRST transformations.  Since the
number of such restrictions is less than the number of coefficients in
the double-copy ans\"atze, one is free to impose some further restrictions
by requiring the expressions for the source currents $j(\varphi)$, 
$j_{\mu\nu}(h)$ and $j_{\mu\nu}(B)$ to take simplified forms.  For
example, one natural such choice could be to require that the
terms appearing on the right-hand sides of eqns (\ref{jphi}), (\ref{hj2})
and (\ref{Bj2}) should involve only the operator $\ibox$ itself, and 
not the higher inverse powers $\iboxsq$ or $\iboxcub$.  

This can indeed be done.  Ostensibly, achieving this would require imposing
a total of seven conditions on the remaining six free parameters in the 
double copy ans\"atze.  However, after imposing the conditions from 
the BRST and anti-BRST transformations, as we did in section 
\ref{sec:BRSTdc}, it turns out that requiring that the five
constants $e_1$, $e_2$, $e_3$, $e_4$ and $e_5$ in eqns (\ref{edefs})
should vanish imposes only four conditions on the remaining 
free parameters.  Thus we obtain a unique solution, which is given by
\bea
a_1 &=& 0\,,\qquad a_2=-\fft{\xi+2}{2\xi}\,,\qquad a_3=-\ft12\,,\qquad 
 a_4= -\fft{1+\xi}{\xi}\,,\nn\\
b_1 &=& \fft1{2-D}\,,\qquad b_2= \fft{1}{(2-D)\, \xi}\,,\qquad
b_3=\fft{1-\xi^2}{(2-D)\,\xi^2}\,,\nn\\
\alpha_1 &=& \fft1{\xi}\,,\qquad \alpha_2= -1 +\fft{1}{\xi^2}\,,\qquad
\gamma=-\ft12\,.\label{iboxonly}
\eea
With these choices, the source currents in (\ref{jphi}), (\ref{hj2})
and (\ref{Bj2}) become simply
\bea
j(\varphi) &=& \ibox\, j^\rho\circ\tilde j_\rho +
  \alpha_1\, \ibox\, j^\alpha(c)\circ \tilde j_\alpha(\tc)\,,\nn\\
j_{\mu\nu}(h)&=& \ibox\, j_{(\mu}\circ \tilde j_{\nu)} +
   b_2\, \eta_{\mu\nu}\, \ibox\, j^\alpha(c)\circ\tilde j_\alpha(\tc)
+ b_1\, \eta_{\mu\nu}\,\ibox\, j^\rho\circ \tilde j_\rho\,,\nn\\
j_{\mu\nu}(B)&=& \ibox\, j_{[\mu}\circ \tilde j_{\nu]}\,.
\eea

  In \cite{duffbrst}, different choices were made for the parameters in the 
convolution product ansatz (\ref{hans}) for $h_{\mu\nu}$.  Specifically,
in \cite{duffbrst} the structures involving $\del j$ or $\del\tilde j$ in
$j_{\mu\nu}(h)$ given in eqn (\ref{hj2}) were required to be absent,
together with requiring $a_4=0$.  After imposing also the conditions
for BRST and anti-BRST invariance, this leads to the conditions given in
eqns (18a), (18b) and (18c) in \cite{duffbrst}, namely
\bea
a_1 &=& \fft1{1-\xi} \,,\qquad a_2=\fft{1+\xi}{2(1-\xi)}\,,
\qquad a_3=-\ft12\,,\qquad
 a_4= 0\,,\nn\\
b_1 &=& -\fft{\xi}{(2-D)(1-\xi)}\,,\qquad 
b_2= -\fft{1}{(2-D)(1-\xi)}\,,\qquad
b_3= -\fft{(1+\xi)}{(2-D)\,\xi}\,.
\eea
With this choice for the parameters the ansatz for $h_{\mu\nu}$ becomes singular
if the gauge parameter $\xi=1$.  There is no such singularity for other
choices, such as in eqn (\ref{iboxonly}).

  We shall consider yet another convenient choice for the parameters 
in section \ref{sec:BH}, when we construct the Schwarzschild
black hole and black string as convolutional double copies.

\section{Black Hole and Black String from Yang-Mills} \label{sec:BH}

In this section we show how the Schwarzschild black hole and the 
black string solution can be obtained explicitly, at the linearised level,
from convolution products of solutions of two Yang-Mills theories.  
In order to do this, it is useful first to construct the explicit
expression for the energy-momentum tensor $T_{\mu\nu}$ in terms of the
source current $j_{\mu\nu}(h)$ that we found in eqn (\ref{hj2}).  

  The equation of motion (\ref{heom}) for $h_{\mu\nu}$ was obtained by 
varying the Lagrangian ${\cal L}'$ given by eqns (\ref{Dlag}), (\ref{hphiBlag}) and (\ref{Dhatlag}), subject also to the condition (\ref{mxi}), and then subtracting out the trace.  Prior to subtracting
the trace, the equation of motion takes the form\footnote{For convenience, we set Newton's constant $G=1$.}
\bea
&&\square  h_{\mu\nu} -\fft{\xi_h +2}{\xi_h}\, (2\del^\rho\del_{(\mu}
h_{\nu)\rho} -\del_\mu \del_\nu h)+ \nn\\
&&
\fft{\xi_h+2}{\xi_h}\, \del^\rho\del^\sigma h_{\rho\sigma}\, \eta_{\mu\nu}
-\fft{\xi_h+1}{\xi_h}\,\square h\, \eta_{\mu\nu} = -16\pi\, T_{\mu\nu}\,,
\eea
where $T_{\mu\nu}$ is the energy-momentum tensor.  Taking out the trace,
and comparing with eqn (\ref{heom}), we see that $T_{\mu\nu}$ is given in
terms of the source current $j_{\mu\nu}(h)$ by
\bea
T_{\mu\nu}=-\fft1{16\pi}\, \big( j_{\mu\nu}(h) -\ft12 \eta^{\rho\sigma}\,
j_{\rho\sigma}(h)\, \eta_{\mu\nu}\big)\,.
\eea
 From eqn (\ref{hj2}) we therefore have that
\bea
T_{\mu\nu} &=& -\fft1{16\pi}\, \Big\{
\ibox\, j_{(\mu}\circ \tilde j_{\nu)} + 
 h_1\, \iboxsq\,\del_\mu\del_\nu\, (j^\rho \circ\tilde j_\rho) +
  h_2\, \iboxsq\, \Big(\del j\circ \del_{(\mu} \tilde j_{\nu)} +
   \del_{(\mu} j_{\nu)}\circ \del\tilde j\Big) \nn\\
&&\qquad\qquad
+h_3\, \iboxcub\, \del_\mu\del_\nu\, (\del j\circ\del\tilde j)
+ h_4\, \iboxsq\,\del_\mu\del_\nu\,
   (j^\alpha(c)\circ\tilde j_\alpha(\tc)) +
h_5\, \eta_{\mu\nu}\,\ibox\, j^\rho\circ \tilde j_\rho\nn\\
&& \qquad\qquad 
+h_6\, \eta_{\mu\nu}\, \iboxsq\,(\del j\circ\del\tilde j) +
h_7\, \eta_{\mu\nu}\, \ibox\, j^\alpha(c)\circ\tilde j_\alpha(\tc)
\Big\}\,,\label{Tmunu}
\eea
where
\bea
h_1 &=& e_2\,,\qquad h_2= -e_1\, \xi\,,\qquad 
 h_3= 2 e_1\, \xi(\xi+1) + e_2\, (\xi^2-1) + e_3\, \xi^2\,,\nn\\
h_4&=&e_5\,,\qquad h_5= (1-\ft12 D)\, b_1 -\ft12 e_2 -\ft12\,,\nn\\
h_6 &=& -e_1\, \xi^2 -\ft12 e_2\, (\xi^2-1) -\ft12 e_3\, \xi^2 +
  (1-\ft12 D)\, e_4\,\xi^2\,,\quad
h_7 = (1-\ft12 D)\, b_2 -\ft12 e_5\,.\label{hi}
\eea

   We shall find it convenient, for what follows, to make a choice for
the $a_i$ and $b_i$ parameters in the ansatz (\ref{hans}) for 
$h_{\mu\nu}$, which are related to $e_i$ via \eqref{edefs}, 
such that $T_{\mu\nu}$ in eqn (\ref{Tmunu}) has a
form that is well adapted to the construction of the linearised
Schwarzschild black hole and black string.  Specifically, we shall
choose the free parameters in the 
convolutional double-copy ans\"atze so that
\bea
h_1=h_2=h_4=h_5=h_7=0
\eea
in the expression (\ref{Tmunu}) for the energy-momentum tensor $T_{\mu\nu}$.
We must also satisfy the equations (\ref{eq3}), (\ref{eq4}) and (\ref{eq5})
that came from requiring BRST and anti-BRST invariance.  We find that a
unique solution exists, with
\bea
a_1=a_2=a_4=0\,,\qquad a_3=-\ft12\,,\qquad
b_1=-b_3=\fft1{2-D}\,,\qquad b_2=0\,.\label{abchoice}
\eea
The remaining non-vanishing $h_i$ coefficients are then given by
\bea
h_3= -1\,,\qquad h_6=0\,.
\eea
To summarise, with the $a_i$ and $b_i$ coefficients in the double-copy
ansatz (\ref{hans}) for $h_{\mu\nu}$ being chosen as in eqn (\ref{abchoice}),
the energy-momentum tensor $T_{\mu\nu}$ is given in terms of the
Yang-Mills source currents by
\bea
T_{\mu\nu} &=& -\fft1{16\pi}\, \big[
\ibox\, j_{(\mu}\circ \tilde j_{\nu)} 
- \iboxcub\, \del_\mu\del_\nu\, (\del j\circ\del\tilde j) 
\big] \,,\label{Tmunu2}
\eea

For the Kalb-Ramond field $B_{\mu\nu}$, we see from the expression
for $j_{\mu\nu}(B)$ in eqn (\ref{Bj1}) that if we choose the
free parameter $\gamma$ to be equal to $-\ft12$, then 
\bea
j_{\mu\nu}(B)&=& \ibox\, j_{[\mu}\circ \tilde j_{\nu]}\,.\label{jmunu(B)}
\eea

  For the dilaton, with $\alpha_2$ determined by the condition of BRST
invariance in eqn (\ref{Qphi}), we can view $\alpha_1$ as a free parameter.
We shall then choose $\alpha_1=1/\xi$ for convenience, as we did in eqns 
(\ref{iboxonly}), and so
\bea
j(\varphi) = \ibox\, j^\rho\circ\tilde j_\rho + \fft1{\xi}\,
 \ibox\, j^\alpha(c)\circ \tilde j_\alpha(\tc)\,.\label{jphi4}
\eea

\subsection{Schwarzschild black hole}\label{sec:schwarzschildbh}

  For definiteness, we shall consider the four-dimensional 
Schwarzschild black hole, but the discussion generalises immediately
to the more general case of the Tangherlini black hole in arbitrary
dimension.  Starting from the standard Schwarzschild form of the metric,
\bea
ds^2 = -\Big(1-\fft{2M}{r}\Big)\, dt^2 + \Big(1-\fft{2M}{r}\Big)^{-1}\, dr^2
+ r^2\, d\Omega_2^2\,,
\eea
it can be recast in the isotropic form by defining a new radial variable $\rho$
such that
\bea
r =\rho\, \Big(1+ \fft{M}{2\rho}\Big)^2\,.
\eea
The Schwarzschild metric then becomes
\bea
ds^2= -\fft{\Big(1-\fft{M}{2\rho}\Big)^2}{\Big(1+\fft{M}{2\rho}\Big)^2}\,
dt^2 + \Big(1+\fft{M}{2\rho}\Big)^4\, dx^i dx^i\,,\label{isotropic}
\eea
where $x^i=\rho\, n^i$ and $n^i$ is a unit vector in the Cartesian 3-space,
with $dn^i dn^i=d\Omega_2^2$.  To linear order in $M$ the metric  
(\ref{isotropic}) takes the form $g_{\mu\nu}=\eta_{\mu\nu} + h_{\mu\nu}$
with $h_{\mu\nu}$ having the non-vanishing components
\bea
h_{00}= \fft{2M}{\rho}\,,\qquad h_{ij}= \fft{2M}{\rho}\, \delta_{ij}\,.
\label{hMNSchw}
\eea
The four Minkowskian coordinates are $x^\mu=(t,x^i)=(t,\bx)$.
It follows from eqns (\ref{hMNSchw}) 
that $\bar h_{\mu\nu}\equiv  h_{\mu\nu}-\ft12 h\, \eta_{\mu\nu}$ is given by
\bea
\bar h_{00}= \fft{4M}{\rho}\,,\qquad \bar h_{ij}=0\,,\qquad
  \bar h_{0i}=0\,.
\eea
In isotropic coordinates the linearised Schwarzschild metric therefore
obeys the De Donder gauge condition $\del^\mu \bar h_{\mu\nu}=0$.  From
the linearised Einstein equation 
$\square \,\bar h_{\mu\nu}=-16 \pi \, T_{\mu\nu}$, we can read off, bearing in
mind that $\square \,\fft1{\rho} = \del_i\del_i\, \fft1{\rho}= -4\pi\, 
\delta^{(3)}(\bx)$ where $\rho= |\bx|$, 
that the energy-momentum tensor that acts as the
source for the Schwarzschild solution has only a $T_{00}$ component,
given by
\bea
T_{00}= M\, \delta^{(3)}(\bx)\,.\label{schwTmunu}
\eea
  
   It is now straightforward to see that if we choose Yang-Mills source
currents to lie in a $U(1)$ subgroup of the gauge group, and to
take the form
\bea
j_\mu = \big(\square \delta^{(4)}(x),0,0,0\big)\,,
\qquad \tilde j_\mu = 
  -16\pi M\, \big(\delta^{(3)}(\bx),0,0,0\big)\,,\label{YMsources}
\eea
then since $\del^\mu \tilde j_\mu=0$, only the first term in the expression
(\ref{Tmunu2}) for the energy-momentum tensor will survive, and we shall
indeed have
\bea
T_{\mu\nu} = M \,  \delta^0_\mu\, \delta^0_\nu\, \delta^{(3)}(\bx)\,,
\eea
which is of precisely the correct form for the energy-momentum 
tensor (\ref{schwTmunu}) for a point-mass located at the 
origin.
Note that the tilded Yang-Mills current
is of the form that corresponds to a point charge of strength $16\pi M$.  The 
untilded Yang-Mills current is just the D'Alembertian acting on 
the four-dimensional delta function $\delta^{(4)}(x)$.  As noted in
appendix A, $\delta^{(4)}(x)$ is the identity for convolution
products, and the D'Alembertian just serves to cancel the $\ibox$ 
occuring in the expression for $T_{\mu\nu}$ in terms of the Yang-Mills
currents. 

  The Kalb-Ramond source current given in eqn (\ref{jmunu(B)}) will 
indeed vanish for the choice of Yang-Mills source
currents in eqns (\ref{YMsources}).  

  For the dilaton, if we take
\bea
j^\alpha(c) = (\iota,0)\, \square \delta^4(x)\,,\qquad 
\tilde j_\alpha(\tc) = -16\pi \xi M  (\bar\iota,0)\,\delta^{(3)}(\bx)\,,
\eea
where $\iota$ and $\bar\iota$ are Grassmann-valued constants with
$\iota\bar\iota$ giving 1, 
then the
contribution from the $c$ and $\tc$ Yang-Mills ghost currents will
cancel the contribution from the Yang-Mills gauge currents in eqn 
(\ref{jphi4}), and so we can arrange for $j(\varphi)$ to vanish.

  In summary, we have seen that we can choose the Yang-Mills sources so 
that the energy-momentum tensor for $h_{\mu\nu}$ has the appropriate form
for a point-mass source for the linearised Schwarzschild solution, 
while there will be no sources for $B_{\mu\nu}$ or $\varphi$, and
so these fields will vanish.

\subsection{Ten-dimensional black string solution}

The ten-dimensional solution describing a black string centred on the
origin of the eight-dimensional transverse space was constructed
in \cite{horostrom}.  In the notation we shall use, it
is given by \cite{dulupo}
\bea
ds^2 &=& W^{-\fft34}\,(-f dt^2+ dx^2) + W^{\fft14}\, (f^{-1}\, dr^2 +
  r^2\, d\Omega_7^2)\,,\nn\\[4mm]
W&=& 1 + \fft{k\sinh^2\delta}{r^6}\,,\qquad 
f = 1 -\fft{k}{r^6}\,,\qquad \varphi= -\ft12 \log W\,,\nn\\
H &=& \lambda \,e^{\varphi}\, 
  {*\omega_7}\,,\qquad \lambda = 3 k\, \sinh2\delta\,,
\label{bstringsol}
\eea
where $*$ denotes the ten-dimensional Hodge dual, and $\omega_7$
denotes the volume form on the unit 7-sphere.  The constants $k$ and
$\delta$ characterise the mass per unit length and the 3-form charge
of the string.

Introducing asymptotically-Minkowskian coordinates $X^M$ with
\bea
X^0=t\,,\qquad X^1= x\,,\qquad X^I= y^I\,,\quad 2\le I\le 9\,,
\eea
where $y^I= r\, n^I$ with $n^I n^I=1$ 
and $dn^I dn^I=d\Omega_7^2$, 
we can write the black string metric up to linear order in $k$ as
$g_{MN}=\eta_{MN} + h_{MN}$, where the non-zero components of $h_{MN}$
are given by
\bea
h_{00}=\fft{k}{r^6}\, \Big(1+ \fft{3 s^2}{4}\Big)\,,\qquad
h_{11}= -\fft{3 s^2}{4 r^6}\,,\qquad
h_{IJ}= \fft{k s^2}{4 r^6}\, \delta_{IJ} + \fft{k}{r^8}\, y^I y^J\,,
\label{hMNstring}
\eea
and we are employing the abreviations $s$ and $c$, where\footnote{There 
should be no confusion between $c=\cosh\delta$ and the ghost $c$
in the untilded Yang-Mills sector of the double-copy construction.}
\bea
s=\sinh\delta\,,\qquad c=\cosh\delta\,.
\eea
 From eqns (\ref{hMNstring}) 
it follows that $\bar h_{MN}\equiv h_{MN} -\ft12 h\, \eta_{MN}$
is given by
\bea
\bar h_{00}=\fft{k c^2}{r^6}\,,\qquad \bar h_{11}= -\fft{k s^2}{r^6}\,,\qquad
\bar h_{IJ}= \fft{k}{r^8}\, y^I y^J\,.\label{barhstring0}
\eea

  Note that $\bar h_{MN}$ given by eqns (\ref{barhstring0}) does not satisfy
the De Donder gauge condition $\del^M\bar h_{MN}=0$.  We can easily perform
a coordinate transformation that puts it in De Donder gauge, at the 
linear order to which we are working, by means of a diffeomorphism
$\xi^M$ such that $\delta h_{MN}= \del_M \xi_N + \del_N \xi_M$.  Specifically,
if we choose the diffeomorphism parameter to be given by
\bea
\xi^0 =0\,,\qquad \xi^1=0\,,\qquad \xi^I= \fft{k\, y^I}{12 r^6}\,,
\eea
then the gauge-transformed $\bar h_{MN}$ will have components given by
\bea
\bar h_{00}=\fft{k}{r^6}\,\Big(c^2 + \fft16\Big)\,,\qquad 
\bar h_{11}= -\fft{k}{r^6}\,\Big(s^2 + \fft16\Big)\,,\qquad
\qquad \bar h_{IJ}=0\,,\
\eea
and all off-diagonal components zero too.  It is easily seen that this
$\bar h_{MN}$ tensor {\it is} in De Donder gauge.  Since we have
\bea
\square \,\fft1{r^6}= \del_I\del_I \,\fft1{r^6} =
 -\del_I \Big(\fft{6 y^I}{r^8}\Big)\,,
\eea
then by integrating over the interior of a sphere centred on the origin
we see that
\bea
\square \, \fft1{r^6} = -6\ \Omega_7\, \delta^{(8)}(\by)\,,
\eea
where $\Omega_7=\ft13 \pi^4$ is the volume of the unit 7-sphere.
It follows that at
the linearised order the energy-momentum tensor, which is related to
$h_{MN}$ by $\square \bar h_{MN}= -16\pi\, T_{MN}$ in De Donder gauge, is
given by
\bea
T_{00}= \fft{3 k \Omega_7}{8\pi}\, \Big(c^2 + \fft16\Big)\,
  \delta^{(8)}(\by)\,,\qquad
T_{11}=- \fft{3 k \Omega_7\, s^2}{8\pi}\, \Big(s^2 + \fft16\Big)\,
  \delta^{(8)}(\by)\,, \label{stringTmunu}
\eea
with all other components vanishing.

   For the construction of this black string solution in terms of the
convolution double copy we shall employ Yang-Mills fields in an
$SU(2)$ subgroup of the gauge group, rather than the $U(1)$ subgroup 
that sufficed for the previous black hole example.\footnote{We could
equally well choose a $U(1)^3$ subgroup, since this would make no
difference at the linearised order at which we are working.}  
We therefore now
denote the Yang-Mills currents by $j^\aa_M$ and $\tilde j^\aa_M$,
where the index {\small $a$} 
will range over 1, 2 and 3.  (We enclose the index
in parentheses to avoid the risk of ambiguities when we assign explicit
numerical values to $a$ in what will follow.)

   It can now be seen that if we choose the Yang-Mills currents
$j^\aa_M$ and $\tilde j^\aa_M$ in the convolutional double copy to be
given by
\bea
j^\sone_M &=& \big(1,1,0,\ldots,0\big)\, \square\,\delta^{(10)}(X)\,,\qquad
\tilde j^\sone_M= -\fft{k \Omega_7}{2}\,  \big(1,-1,0,\ldots,0\big)\,
 \delta^{(8)}(\by)\,,\nn\\
j^\stwo_M &=& \big(1,-1,0,\ldots,0\big)\, \square\,\delta^{(10)}(X)\,,\qquad
\tilde j^\stwo_M= -\fft{k \Omega_7}{2}\,  \big(1,1,0,\ldots,0\big)\,
 \delta^{(8)}(\by)\,,\nn\\
j^\sthree_M &=& \big(c,s,0,\ldots,0\big)\, \square\,\delta^{(10)}(X)\,,\qquad
\tilde j^\sthree_M= -6k\,\Omega_7\,  \big(c,-s,0,\ldots,0\big)\,
 \delta^{(8)}(\by)\,,
\label{YMstringsources}
\eea
then for the parameter choices in which the energy-momentum tensor
is given by eqn (\ref{Tmunu2}), we see that $T_{\mu\nu}$ in eqn (\ref{Tmunu2})
is precisely equal to the energy-momentum tensor we found in 
eqn (\ref{stringTmunu}) for the black string.  The reason for needing
both the $a=1$ and $a=2$ currents, which contribute equally to
$T_{MN}$, will be seen below when we consider the source current
$J_{MN}(B)$ for the Kalb-Ramond field.
 
  It can be seen from the expression for the 3-form field strength $H$ in
eqns (\ref{bstringsol}) that one can write
the 2-form potential $B$ in a gauge where it is given just by
\bea
B_{01} = -\fft{k s c}{W\, r^6} = -\fft{k s c}{r^6} + {\cal O}(k^2)\,.
\eea
Clearly this is written in a Lorenz-like gauge where $\del^M B_{MN}=0$.  
At the linear order in $k$ it
obeys $\square B_{01}= 6 k \Omega_7\, s c\, \delta^{(8)}(\by)$.  
Comparing with eqn (\ref{Beom}) we see that the source current $j_{MN}(B)$
is given by
\bea
j_{01}(B)=  6k \Omega_7\, s c\, \delta^{(8)}(\by)\,.
\eea
It can be seen that this is precisely consistent with what we obtain by
substituting the Yang-Mills source currents of eqns (\ref{YMstringsources}) 
into eqn (\ref{Bj2}).  Note that the $a=1$ Yang-Mills currents 
would contribute an additional, unwanted, term to $j_{MN}(B)$, which is
cancelled by an equal but opposite contribution from the $a=2$ currents.

  The dilaton in the solution in eqns (\ref{bstringsol}) is given by
\bea
\varphi = -\fft{k s^2}{2 r^6}
\eea
at the linear order in $k$, and so from (\ref{phieom}) we find that the
dilaton current is given by
\bea
j(\varphi) = 3 k\, \Omega_7\, s^2\, \delta^{(8)}(\by)\,.
\eea
We see from eqns (\ref{YMstringsources}) and (\ref{jphi4}) that we can 
obtain this current in the double copy construction by choosing the Yang-Mills
ghost currents
\bea
j^\alpha(c) = (\iota,0)\, \square \delta^{(10)}(X)\,,\qquad 
  \tilde j_\alpha(\tc) = \mu\, (\bar\iota,0)\, \delta^{(8)}(\by)\,,
\label{stringghostsources}
\eea
with $\iota\bar\iota$ giving 1 and the constant $\mu$ chosen to be
\bea
\mu= \xi\, k\, \Omega_7\, (1-9c^2)\,.\label{mustring}
\eea

\subsection{BPS string solution}

   It is interesting also to consider the BPS limit of the black string,
which is obtained by sending the charge parameter $\delta$ to infinity
while simultaneously sending $k$ to zero, holding the product
$q\equiv k\sinh^2\delta$ fixed.  Thus in the BPS limit one has
\bea
k s^2\longrightarrow q\,,\qquad k c^2\longrightarrow q\,.
\eea
It can be seen from eqn (\ref{stringTmunu}) that the
energy-momentum tensor for the BPS string is given by
\bea
T_{00}= \fft{3 q \Omega_7}{8\pi}\,
  \delta^{(8)}(\by)\,,\qquad
T_{11}=- \fft{3 q \Omega_7}{8\pi}\,
  \delta^{(8)}(\by)\,, \label{BPSstringTmunu}
\eea
with all other components vanishing.
Since one can always make a see-saw rescaling of the expressions for
the untilded and tilded Yang-Mills currents in (\ref{YMstringsources}),
of the form $j^\aa_M\longrightarrow k^{1/2}\, j^\aa_M$ and 
$\tilde j^\aa_M\longrightarrow k^{-1/2}\, \tilde j^\aa_M$, it is
easily seen that the BPS string solution can be obtained as the
convolutional double copy of just $U(1)$ Yang-Mills sources, with
\bea
j^\sthree_M = \big(1,1,0,\ldots,0\big)\, \square\,\delta^{(10)}(X)\,,\qquad
\tilde j^\sthree_M= -6q\,\Omega_7\,  \big(1,-1,0,\ldots,0\big)\,
 \delta^{(8)}(\by) \,.\label{BPSstring}
\eea
The ghost currents can again be chosen as in eqn (\ref{stringghostsources}),
with the constant $\mu$, following from eqn (\ref{mustring}), given by
\bea
\mu= -9\xi\, q\, \Omega_7\,.
\eea

\section{Discussion and Conclusions}\label{sec:conc}

  In this paper, we have explored in greater detail some of the ideas
presented in \cite{duffbrst} for obtaining ``${\cal N}=0$ supergravity''
(i.e.~the low-energy limit of the bosonic string) as a convolutional
double copy of Yang-Mill squared.  A crucial aspect of the construction in
\cite{duffbrst} is the inclusion of the ghosts and antighosts of a BRST
description, in order to capture fully the degrees of freedom of
the theory.  Accordingly, our discussion has been centred around the
BRST description of the theory and its gauge invariances, with an 
emphasis on the anti-BRST as well as the BRST symmetries of the system.

   We have illustrated the application of the convolutional 
double copy in two examples.  The first, which has been discussed 
previously, is the construction of the four-dimensional 
static black hole solution
of ${\cal N}=0$ supergravity (i.e.\ the Schwarzschild black hole)
as a double copy of two Yang-Mills solutions, one of which describes
a $U(1)$ static point charge and the other a $U(1)$ configuration 
corresponding to a $\square\delta^{(4)}(x)$ source.  Our second example,
which is more involved, is the ten-dimensional black string solution
of ${\cal N}=0$ supergravity.  We showed how this can be constructed 
as a double copy involving non-abelian $SU(2)$ Yang-Mills solutions in
the two copies.  As far as we are aware, this is the first example that
has been obtained in the literature in which multi-component Yang-Mills
configurations are necessary in order to describe a ${\cal N}=0$
supergravity solution.  We showed also how the BPS string solution
arises as a limit in which the $SU(2)$ Yang-Mills reduces to $U(1)$.

   All of the discussion in our paper has been, as in \cite{duffbrst}, 
at the linearised level.  There have been some discussions that 
push beyond the linear level, such as the construction up to cubic
order in the Lagrangian, in \cite{Borsten:2020xbt}.  It would be of
interest to pursue these investigations further.

\section*{Acknowledgements}

We thank Michael Duff, Yue-Zhou Li, Noah Miller, Ricardo Monteiro, Malcolm Perry and Junchen Rong for useful discussions. MG and CNP would like to thank the Albert
Einstein Institute, Potsdam, for hospitality during the course of this work. MG is supported by a Royal Society 
University Research Fellowship. CNP is partially supported by DOE grant DE-FG02-13ER42020.

\appendix

\section{Convolution Product and Convolution Inverse}

\subsection{Basic properties of the convolution}

 The convolution is defined, for functions $f(x)$ and $g(x)$ in a
$D$-dimensional spacetime, by
\bea
(f\star g)(x) = \int d^D y\, f(x-y)\, g(y)\,.
\eea
The convolution product is associative, so 
$f\star (g\star h)=(f\star g)\star h$.
For functions obeying appropriate boundary conditions, one has
\bea
\del_\mu(f\star g)= (\del_\mu f)\star g= f\star (\del_\mu g)\,.
\eea
If $\cF_k$ denotes the Fourier transform with respect
to the D-momentum $k$, so that
\bea
\cF_k(f) \equiv \int d^D x\, f(x)\, e^{-\im k\cdot x}\,,
\eea
then 
\bea
\cF_k(f\star g) = \cF_k(f)\, \cF_k(g)\,.\label{conth}
\eea

The convolution inverse $\tilde f(x)$ of a function $f(x)$ is defined by
\bea
(f\star \tilde f)(x)=\delta^D(x)\,.\label{CIdef}
\eea
Note that the function $f(x)= \delta^D(x)$ is equal to its own 
convolution inverse.  In other words, $\delta^D(x)$ is the identity in
the convolution product.  At the level of Fourier transforms, one can see
from (\ref{conth}) that the
convolution inverse $\tilde f$ and the original function $f$ are such that
\bea
\cF_k(\tilde f)= \fft1{\cF_k(f)}\,.\label{FourierCI}
\eea

\subsection{Non-existence of convolution inverse of $\fft1{r}$}

   One might think that a natural expectation for the Yang-Mills 
configurations that would give rise to the Schwarzschild black hole as
the convolutional double copy would be to take point-particle solutions
for the left and right Yang-Mills potentials, as, for example, in
\cite{point}. Taking four dimensions 
for simplicity, the
Yang-Mills potentials for point charge particles would be of the form
$A^a_0\sim \fft{e}{r}$, and $\wtd A^{\tilde a}_0\sim \fft{\tilde e}{r}$.  
In order
to get a linearised metric perturbation of the form $h_{00}\sim \fft{m}{r}$,
this would then require that the spectator field $\phi^{-1}_{a\tilde b}$
in eqn (\ref{convprod}) should have the general form of the convolution
inverse of $\fft1{r}$.  However, it can easily be seen that the convolution
inverse of $\fft1{r}$ does not exist.  This can be seen in terms of the
Fourier transform, as follows.  The Fourier transform of $\fft1{r}$ is given
by 
\bea
\cF_k\Big(\fft1{r}\Big) = \fft{\delta(k^0)}{{\bf k}^2}\,,
\eea
and so it follows from eqn (\ref{FourierCI}) that the Fourier transform of
the convolution inverse of $\fft1{r}$ would be
\bea
\fft{{\bf k}^2}{\delta(k^0)}\,.
\eea
This is a function that is ``infinite almost everywhere,'' and thus does
not really make sense.  Thus while it is true that one could write a
formal expression for the Fourier transform of the convolution products
of two $\fft1r$ functions with a ``convolution inverse'' of $\fft1r$ as
\bea
\fft{\delta(k^0)}{{\bf k}^2} \times \fft{{\bf k}^2}{\delta(k^0)} \times
 \fft{\delta(k^0)}{{\bf k}^2} = \fft{\delta(k^0)}{{\bf k}^2}\,,
\eea
conveying the notion that one would thereby obtain a result of $\fft1r$ 
for the linearised metric contribution, this seems to be more akin to
a mnemonic that lacks mathematical justification.

   The non-existence of the convolution inverse of $\fft1{r}$ can also be
seen directly in position space.  Adopting the notation that the position
4-vector $x^\mu$ has components $(x^0,{\bf x})$, 
we see from eqn (\ref{CIdef}) 
that the convolution inverse of the function $\fft1{r}$
would be a function $\tilde f(x)$ such that
\bea
\int d^4y\, \fft{\tilde f(y)}{|{\bf x}-{\bf y}|} &=& \delta^4(x)\nn\\
&=& \delta(x^0)\, \delta^3({\bf x})\,.
\eea
The integral on the left-hand side is independent of $x^0$.  However, the
delta function on the right-hand side includes a factor $\delta(x^0)$.  Thus,
there cannot exist any function $\tilde f(x)$ such that it is the
convolution inverse of $f(x)=\fft1{|{\bf x}|}$.

\subsection{Non-existence of convolution pseudo-inverse of $\fft1{r}$}

The 
convolution pseudo-inverse $F$ of a function $f$ can be defined 
by the equation
\bea
f\star F\star f = f\,.
\eea
Writing this out explicitly, this means
\bea
\int d^D y\, \int d^4 z f(x-y-z)\, F(z) f(y) = f(x)\,.
\eea
If we consider the previous four-dimensional example, and now try
to calculate the convolution pseudo-inverse $F(x)$ 
of the function $f(x)=\fft1{r}$,
where $r=|\bx|$, we would then require that
\bea
\int d^4y\, \int d^4 z  \fft1{|{\bf x} - {\bf y} -{\bf z}|}\, \fft1{|{\bf y}|}\,
F(z) = f(x)\,.
\eea
The $y$ integration does not
involve dependence on 
the unknown function $F(z)$ that we are trying to solve for,
and so for this we may first consider just the $y$ integration, defining
\bea
G(x-z)\equiv \int d^4y\,\fft1{|{\bf x} - {\bf y} -{\bf z}|}\,\fft1{|{\bf y}|}\,.
\eea
There are two convergence problems here.  First of all, the integrand
does not depend on $y^0$, and so the $y^0$ integration will give 
infinity.  Over and above that, the remaining spatial integrations will
be of the form
\bea
\int d^3{\bf y}\, \fft1{|{\bf w} - {\bf y}|}\, \fft1{|{\bf y}|}\,,
\eea
where ${\bf w}= {\bf x}-{\bf z}$.  This integration will diverge at
large $|{\bf y}|$. 

In short, the ``convolutive pseudo-inverse'' of $1/r$ would appear not to 
exist. In general, it would appear that convolution inverses of functions
that do not depend on all $D$ spacetime coordinates do not exist.

\subsection{Convolutional double copy ansatz using well-defined
convolution inverse}

  For the reasons seen in the previous subsections, it does not seem to
be possible to construct a convolutional double copy for an object such as
a black hole by taking the product of tilded and untilded Yang-Mills
point charges.  
For this reason, the kinds of ans\"atze we have used in this
paper have been along the lines of those employed in 
\cite{Cardoso:2016amd,Cardoso:2016ngt}. 
Such an ansatz is asymmetric, in the sense that one copy of the
Yang-Mills potential is taken to be that of a point charge, while the other is
taken to be of the form of a full $D$-dimensional delta function
$\delta^D(x)$.   The spectator field is taken to be the convolution inverse
of this, and so it exists and is in fact also $\delta^D(x)$.

\bibliographystyle{utphys}
\bibliography{refs}

\end{document}